\documentclass[pra,aps]{revtex4}
\usepackage{graphicx,amsfonts,bm,amsmath}
\newcommand{\kk}{\bm{k}}
\newcommand{\qq}{\bm{q}}
\newcommand{\bq}{b_{\bm{q}}}
\newcommand{\bk}{b_{\bm{k}}}
\newcommand{\bkd}{b_{\bm{k}}^{\dag}}

\newcommand{\sumk}{\sum_{\bm{k}}}
\newcommand{\sumq}{\sum_{\bm{q}}}
\newcommand{\wk}{\omega_{\bm{k}}}
\newcommand{\wq}{\omega_{\bm{k}}}

\newcommand{\gk}{g_{\bm{k}}}
\newcommand{\gq}{g_{\bm{q}}}

\newcommand{\rl}{\rangle\!\langle}
\newcommand{\idw}{\int d\omega}

\DeclareMathOperator{\re}{Re}

\begin{document}
\title{Phonon-assisted excitation transfer in quantum dot molecules}
\author{Emil Rozbicki and Pawe{\l} Machnikowski}
 \affiliation{Institute of Physics, Wroc{\l}aw University of
Technology, 50-370 Wroc{\l}aw, Poland}

\begin{abstract}
\begin{center}
\parbox{145mm}{
We derive a quantum-kinetic description of phonon-assisted F{\"o}rster
transfer between two coupled quantum dots (a quantum dot molecule). 
We show that the exciton state decays to the ground state of the QDM
via a combination of the Rabi rotation and exponential decay. For
moderately spaced dots this process takes place on a picosecond time scale.
}
\end{center}\end{abstract}


\maketitle

Coupled quantum dots (QDs),
referred to as quantum dot molecules (QDMs),
have attracted much attention in recent years
\cite{bayer01-ortner03-krenner05b}. 
Besides the electronic coupling or superradiance effects
\cite{scheibner07-sitek07a}, the properties of QDMs
are affected by phonon-related phenomena. In particular,
many experiments have shown phonon-assisted excitation transfer 
\cite{heitz98-tackeuchi00-ortner05c-nakaoka06} between the QDs. 
For moderate separations between the dots ($\gtrsim 6$~nm) tunneling is exponentially suppressed and
the energetically lowest states correspond to spatially direct excitons
localized in individual QDs \cite{szafran01}. Such states are bound by
the Coulomb interaction via interband dipole moments
\cite{lovett03b-ahn05}, i.e., by the F{\"o}rster
interaction \cite{forster48-dexter53}. 
Signatures of such coupling were indeed found in a
photon-correlation experiment \cite{gerardot05}. 

In this paper we derive a quantum-kinetic description of the
evolution of an exciton in a QDM, including the effect of
the carrier-phonon coupling. We show that, depending on the
parameters, the system 
can show different dynamical scenarios, ranging from partial pure
dephasing to an almost exponential transfer. As we shall see,
the transfer may be very fast (on the time scales of several
picoseconds), that is, 2-3 orders of magnitude faster than suggested
by the existing perturbative estimates \cite{govorov05}.

We consider two flat, cylindrically symmetric, coaxial QDs, separated
by the distance $D$ along the $z$ 
axis and interacting with phonons. The formalism will be restricted to
the subspace, spanned by
the states $|0\rangle,|1\rangle$, corresponding to a single exciton in the
ground state of the lower and upper dot, respectively (with a fixed
polarization). We assume that the wavefunctions of excitons confined
in different dots do not overlap, so that no phonon-assisted
transitions are possible. The Hamiltonian of the system is then
\begin{equation}\label{ham}
H  =  \frac{1}{2}\Delta \sigma_{z}+V\sigma_{x}+\hbar\sumk\wk\bkd\bk
+\sum_{l=0,1}|l\rl l|\sumk\left(\gk^{(l)}\bk+\gk^{(l)*}\bkd\right),
\end{equation}
where $\sigma_{i}$ are Pauli matrices in te basis
($|0\rangle,|1\rangle$), $\Delta$ is the energy mismatch between the dots, 
$V$ is the amplitude of the F{\"o}rster coupling, $\bkd,\bk$ are
creation and annihilation operators for the phonon mode with a wave vector 
$\kk$, $\wk$ is the corresponding frequency and $\gk^{(l)}$ are
exciton-phonon coupling constants. For heavy-hole excitons confined in
QDs stacked along $z$ one has 
$V=d^{2}/(4\pi\epsilon_{0}\epsilon_{\mathrm{r}}D^{3})
=(3/4)[\hbar c/(DE)]^{3}\hbar\Gamma$,
where $d$ is the interband dipole moment, $\epsilon_{0}$ is the vacuum
dielectric constant, $\epsilon_{\mathrm{r}}$ is the relative
dielectric constant of the crystal, $c$ is the speed of light and
$\Gamma$ is the inverse exciton lifetime.

The most effective interaction between neutral excitons and
phonons is the deformation potential coupling to longitudinal acoustic
phonons. Approximating the exciton wave function by a product of
identical Gaussians one gets
\cite{roszak05b}
\begin{displaymath}
\gk^{(0,1)}=(\sigma_{\mathrm{e}}-\sigma_{\mathrm{h}})
\sqrt{\frac{\hbar k}{2\rho v u}}e^{-(lk_{\bot}/2)^{2}-(l_{z}k_{z}/2)^{2}}
e^{\pm ik_{z}D/2},
\end{displaymath}
where $\sigma_{\mathrm{e}},\sigma_{\mathrm{h}}$ are the deformation
potential constants for electrons and holes, $v$ is the
normalization volume for phonons, $k_{\bot},k_{z}$ are the components
of the wave vector in the QD plane and along $z$,
$l_{\bot},l_{z}$ are the confinement sizes in these two directions,
and $u$ is the speed of sound.

The evolution of the interacting carrier-phonon system is described
using the correlation expansion technique 
\cite{rossi02,forstner03-krugel05,krugel06}. 
One starts from the three dynamical
variables $x,y,z$ describing the carrier state, 
$x=\langle\sigma_{x}(t)\rangle,\ldots$, where 
$\sigma_{i}(t)=e^{iHt/\hbar}\sigma_{i}e^{iHt/\hbar}$ are the operators
in the Heisenberg picture. From the Heisenberg equations of motion one
finds the dynamical equations for these three variables,
\begin{equation}
\label{x}
\dot{x}=i\langle[H,\sigma_{x}]\rangle
=-\Delta y-4y\sumk\re B_{\kk}-4y\sumk\re y_{\kk},
\end{equation}
and analogous for $y$ and $z$ (from now on, the time dependence will
not be written explicitly). Obviously, this set of equations is not
closed, but involves the new
phonon variables $B_{\kk}=\gk\langle\bk\rangle$, as well as 
phonon-assisted variables of the form
$y_{\kk}=\gk\langle\langle \sigma_{y}\bk\rangle\rangle=
\langle \sigma_{y}\bk\rangle
-\langle \sigma_{y}\rangle\langle\bk\rangle$. 
Here $\gk=(\gk^{(0)}-\gk^{(1)})/2$ and
the double angular brackets,
$\langle\langle\ldots \rangle\rangle$,  denote the
correlated part of a product of operators, obtained by
substracting all possible factorizations of the product. 

Next, one writes down the equations of motion for the new variables
that appeared in the previous step, for instance,
\begin{eqnarray}\label{yk}
\dot{y}_{\kk} & = & i\langle[H,y_{\kk}]\rangle
=\Delta x_{\kk}-2Vz_{\kk}-i\wk y_{\kk}
+|\gk|^{2}(iyz+x)\\
\nonumber
&&+2\sumq (x_{\qq\kk}+\tilde{x}_{\qq\kk})
+4x_{\kk}\sumq\re B_{\qq}
+2x\sumq (B_{\qq\kk}+\tilde{B}_{\qq\kk}),
\end{eqnarray}
where the new two-phonon and two-phonon-assisted variables are defined as
$B_{\qq\kk}=\gq\gk\langle\langle \bq\bk\rangle\rangle$, 
$\tilde{B}_{\qq\kk}=\gq^{*}\gk\langle\langle
\bq^{\dag}\bk\rangle\rangle$, 
$x_{\qq\kk}=\gq\gk\langle\langle \sigma_{x}\bq\bk\rangle\rangle$,
$\tilde{x}_{\qq\kk}=\gq^{*}\gk\langle\langle
\sigma_{x}\bq^{\dag}\bk\rangle\rangle$, etc. In the next step, one
writes the equation of motion for these new variables, introducing
three-phonon variables. It is clear that the resulting hierarchy of
equations in infinite and has to be truncated at a certain level. Here
we do this by setting all the correlated parts of three-phonon and
three-phonon assisted variables equal to zero. This amounts to
neglecting three-phonon processes (that is, emission or
absorption of three or more phonons within the memory time of the
phonon reservoir, which is of order of 1 ps).
The motivation for this
procedure is that higher order correlations should play a decreasing
role in the dynamics. From the equations of motion it is also clear
that such higher order correlations develop in higher orders with
respect to the coupling constants $\gk$.

In this way we treat the problem at the same level as in the recent
work \cite{krugel06}, taking into account the coherent and
non-equilibrium phonons which are essential for the correct
description of carrier phonon-kitetics in QDs \cite{krugel06}.
As an improvement over the standard approach
\cite{forstner03-krugel05,krugel06}, we define collective,
frequency-dependent variables
\begin{eqnarray*}
B_{1}(\omega) =\sumk\delta(\omega-\wk)B_{\kk},&
B_{2}(\omega,\omega')=\sum_{\qq\kk}
\delta(\omega-\wq)\delta(\omega'-\wk)B_{\qq\kk}, & \\
x_{1}(\omega)=\sumk\delta(\omega-\wk)x_{\kk},&
x_{2}(\omega,\omega')=\sum_{\qq\kk}
\delta(\omega-\wq)\delta(\omega'-\wk)
x_{\qq\kk},&
\mathrm{etc.}
\end{eqnarray*}
All the equations of motion can be rewritten in terms of these
frequency-dependent variables. For instance, Eqs.~(\ref{x}) and
(\ref{yk}) read
\begin{eqnarray*}
\dot{x}& \!=\! & -\Delta y-4y\idw\re B_{1}(\omega)-4\idw\re y_{1}(\omega),\\
\dot{y}_{1}(\omega) & \!=\! & 
  \Delta x_{1}(\omega)-2Vz_{1}(\omega)
  -i\omega y_{1}(\omega)+J(\omega)(iyz+x)
+4x_{1}(\omega)\idw'\re\! B_{1}(\omega') \\
&&+2\idw' \left[x_{2}(\omega',\omega)+\tilde{x}_{2}(\omega',\omega)\right]
+2x\idw'\left[B_{2}(\omega',\omega)+\tilde{B}_{2}(\omega',\omega)\right],
\end{eqnarray*}
where $J(\omega)=\sumk|\gk|^{2}\delta(\omega-\wk)$.
In this way, the set of variables labeled by points in the
3-dimensional reciprocal space is replaced by a set labeled by points
on a real frequency axis. For the calculations, the frequency
axis is discretized and the ordinary differential equation for the
resulting variables is integrated numerically (we use 601 uniformly
spaced points up to the frequency cutoff at 20~ps$^{-1}$, which
yields $5.8\cdot10^{6}$ variables).

\begin{figure}[tb]
\begin{center}
\unitlength 1mm
\begin{picture}(125,65)(0,5)
\put(0,30){\resizebox{125mm}{!}{\includegraphics{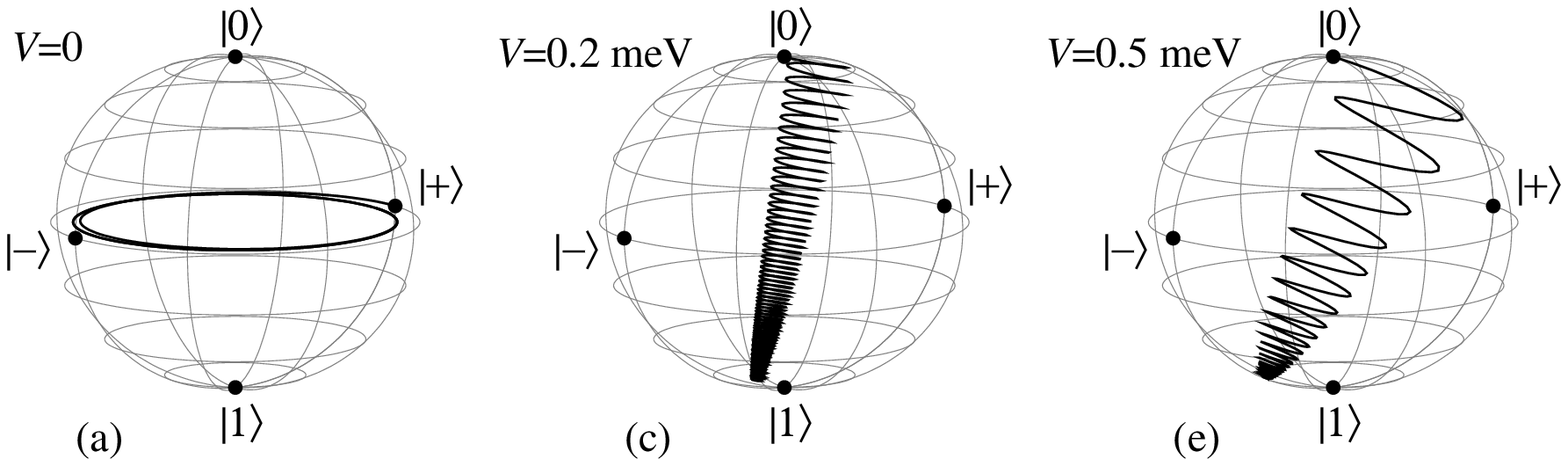}}}
\put(0,0){\resizebox{125mm}{!}{\includegraphics{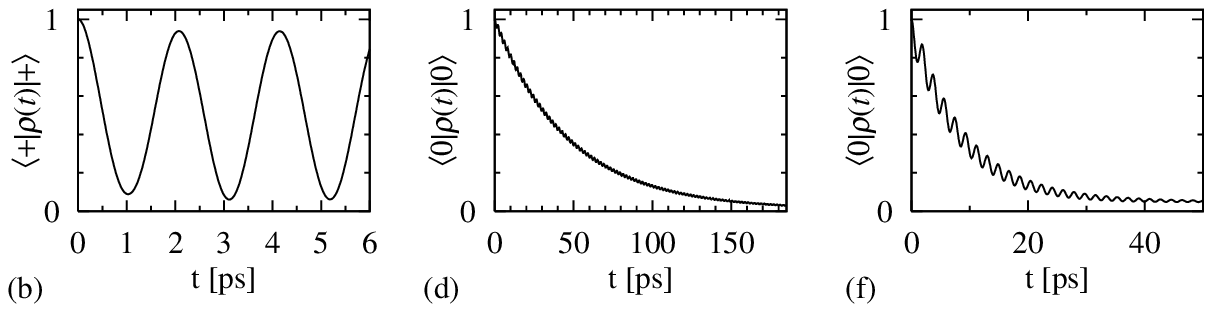}}}
\end{picture}
\end{center}
\caption{(a,c,e) The Bloch sphere representation of the evolution in
the three cases desribed in the text. (b) The overlap of
the system state with the initial state $|+\rangle$. (d,f) The
occupation of the higher energy state as a function of time for
two values of the F{\"o}rster interaction. Here $l_{\bot}=4.5$~nm,
$l_{z}=1$~nm, $D=6$~nm and the temperature $T=0$.}
\label{fig}
\end{figure}

Let us begin the presentation of the results with the noninteracting
case, $V=0$. The occupation of each dot is then conserved, so that no
excitation transfer may take place. Nonetheless, this does not mean
that no carrier-phonon kinetics takes place. In Fig.~\ref{fig} we present the
evolution after an instantaneous preparation of the state 
$|+\rangle=(|0\rangle+|1\rangle)/\sqrt{2}$ in a pair of QDs with
$\Delta=2$~meV. The pure dephasing effect, related to the
lattice response to the apperance of a charge distribution
\cite{machnikowski07a}, reduces the coherence of the superposition
state. This is also manifested by the decreasing amplitude of 
oscillations of the projection on the initial state, 
$\langle +|\rho(t)|+\rangle$, where $\rho(t)$ is the reduced density
matrix of the carrier subsystem.

In the presence of an interaction, $V\neq 0$, the occupations of the
two dots are not conserved and excitation transfer becomes
possible. As can be seen in Figs.~\ref{fig}(c-e), the system evolution
is a combination of a rotation around a tilted axis, defined by the
eigenstates of the unperturbed Hamiltonian, and dephasing resulting
from the interaction with phonons. As a result of the latter, at $T=0$
the system relaxes towards the lower eigenstate of the unperturbed
Hamiltonian (with some correction due to phonon-induced
energy shifts and polaron effects). For $V\ll \Delta$, this lower
eigenstate is close to the state $|1\rangle$.
In this limit, the transfer is nearly exponential, as can be seen
in Fig.~\ref{fig}(d). 

When the interaction gets stronger the final state
gains more admixture of the higher-energy dot. Now, the transfer takes
place via clearly marked oscillations about an exponential trend. The
value of $V=6$~meV used in Figs.~\ref{fig}(e,f) corresponds to the
interband diple moment of $9$~$e${\AA} (or a radiative lifetime of 470
ps) in the dipole approximation, which is within the range of
typical values for self-assembled structures. For this value, the
decay rate is about 10 ps, which means that the process is very fast.

The results presented in this paper show that the phonon-assisted
excitation transfer in QDMs is an efficient process that may
considerably affect the observable properties of these systems. For
moderately spaced dots, the F{\"o}rster coupling results in a fast
transfer 
showing a pattern of oscillations around an exponential
curve. We believe that these characteristic signatures should make it
possible to identify this process in experimental observations.

P.M. is grateful to V.~M.~Axt for many discussions on the
correlation expansion.





\end{document}